\documentclass[twocolumn,superscriptaddress]{revtex4-1}

\usepackage{mathrsfs}
\usepackage{amsmath}
\usepackage{amssymb}
\usepackage{graphicx}
\usepackage{float}
\restylefloat{table}
\usepackage[usenames,dvipsnames,svgnames,table]{xcolor} 
\usepackage[normalem]{ulem}
\newcommand\risco{\bgroup\markoverwith
{\textcolor{red}{\rule[0.4ex]{2pt}{1.2pt}}}\ULon} 
\newcommand\sublin{\bgroup\markoverwith
{\textcolor{blue}{\rule[-0.4ex]{2pt}{1.2pt}}}\ULon} 

\begin{document}

\title{ Oscillating properties of a two-electron quantum dot  in the presence of a magnetic field}
\author{A. M. Maniero}
\email{angelo.maniero@ufob.edu.br}
\affiliation{Universidade Federal do Oeste da Bahia, 47808-021, Barreiras, BA, Brazil}

\author{C. R. de Carvalho}
\email{crenato@if.ufrj.br}
\affiliation{Instituto de F\'{\i}sica, Universidade Federal do Rio de Janeiro, Rio de Janeiro, 21941-972, RJ, Brazil}

\author{F. V. Prudente}
\email{prudente@ufba.br}
\affiliation{Instituto de F\'{\i}sica, Universidade Federal da Bahia, Campus Universit\'ario de Ondina, 40170-115, Salvador, BA, Brazil}

\author{Ginette Jalbert}
\email{ginette@if.ufrj.br}
\affiliation{Instituto de F\'{\i}sica, Universidade Federal do Rio de Janeiro, Rio de Janeiro, 21941-972, RJ, Brazil}

\date{\today }
\begin{abstract}
We study the system consisted of two electrons in a quantum dot with a three-dimensional harmonic confinement potential under the effect of a magnetic field. Specifically, two different confinement conditions are considered, one isotropic three-dimensional and the other  anisotropic $quasi$-two-dimensional. Singlet and triplet  lowest states properties as  the  energy, the exchange coupling, the two-electron density function and the spatial spreading of the two electrons in terms of the variance along the $x$-direction are analyzed. In this study we employ the full configuration interaction method with Cartesian anisotropic Gaussian-type orbitals as basis set. These functions allow to explore the confining characteristics of a potential due to their flexibility of using different exponents for each direction in space. The convergence of the results, depending on the size of the set of basic functions, is examined and the oscillations of  different physical quantities, concerning the singlet and triplet states, as a function of the magnetic field are discussed.

\end{abstract}

\maketitle


\section{\label{sec:I} INTRODUCTION}

In the last three decades the area of nanostructures, an overlapping field of new materials and solid state physics, have experienced a huge development. The advance of semiconductor structure technology at the nanoscale \cite{heinzel2003nanostructures, natelson2015nanostructures} has increased interest in studying the properties of confined quantum systems \cite{sabin2009v57-58}. In this context, one finds that few-electron quantum dots play an important role \cite{kouwenhoven2001,chan2004}. The usefulness of quantum dots (QDs) appears in a wide range of scientific areas spreading from quantum information \cite{mazurek2014,ayachi2014,roszak2015} to biology and medicine \cite{efros2018}. In late 90s one has noticed that the physical properties of QDs are directly connected with their geometry, i.e. their size and form \cite{alivisatos1996nano, brauman1996clusters}, and since early 00s an ordinary approach to treat this issue has been in terms of the profile and strength of the confinement potential \cite{bielinska2001,sako2003}. 

The influence of external fields in the QDs properties was also studied by that time; for instance the behavior of the exchange coupling $J$, as well as the double occupation of a dot by the two electrons of a two-electron double QD  was studied as a function of a static magnetic field and the potential barrier height. The latter being controlled via effective mass with a laser field \cite{carvalho2003}. In the same vein, the dependence of the ground state of one and two electrons, as well as of the exchange energy $J$ (two electrons case), was also investigated for a double QD as a function of an external magnetic field and of the potential barrier between the dots \cite{melnikov2006}. It is worth mentioning that two-electron double QD still remains as a promising hardware to implement quantum computation \cite{ayachi2014, mazurek2014, roszak2015}.
 
We have recently developed a numerical code in order to study confined quantum systems, more specifically, electrons in QDs. In order to test our computational code we have studied the behavior of two electrons within a QD, whose confining potential was a 3D anisotropic parabolic one \cite{olavo2016}. In sequence, by making use of  our computational tools we have addressed the issue of a two-electron double QD under the effect of a laser field \cite{maniero2019}, a subject previously seen in Ref.~\cite{carvalho2003}. However, we have not considered the presence of a static magnetic field, for this leads to a modification in the Hamiltonian that demands a change in the computational code that we have left to treat later. In fact this was the motivation of the present article; to take into account the effect of an external magnetic field. Thus, in order to accomplish this purpose, we have done the corresponding modification in the computational code that, in few words, consisted in using complex coefficients in the linear combination of the basis functions.

Although our focus is to study two-electron double QDs under effect of external influence, we have decided to turn back to a known problem -- two electrons in a single QD  described by 3D parabolic anisotropic confining potential -- as a strategy to deal with the modification done in our code. By doing so, an unexpected behavior of some physical quantities have arisen naturally from our computations; they  present an oscillatory comportment as function of the magnetic field which drew our attention. It was then we became aware of work of de Haas and van Alphen \cite{deHaas1930}. In their work they  investigated the diamagnetism of metals by analyzing the dependence of the susceptibility of metals under a magnetic field. They argued that to correctly understand the phenomenon of diamagnetism one must admit that the conduction electrons are not completely free; they must be in some extension bond in the material, under the influence of neighboring atoms. Therefore, in a certain sense the conduction electrons must be {\it confined} in the material sites. Besides, they were the first to observe experimentally an {\it oscillatory} behavior of the susceptibility as function of the magnetic field which is known in the literature as the {\it de Haas-van Alphen} (dHvA) effect.

The subject of electrons under a magnetic field has been of interest for a long time and several authors have theoretically addressed this issue \cite{Fock1928, Landau1930, Darwin1931, Sondheimer1951, Dingle1952}. In all these treatments the medium has been described as a bunch of free electrons. In particular, in Refs.~\cite{Sondheimer1951, Dingle1952} despite the authors were concerned with the dHvA effect they were unable to carry out the assumption suggested by de Haas and van Alphen that the conducting electrons are not entirely free. More recently, in the framework of QDs, the electron-electron interaction as well as a confining potential have been theoretically taken into account and an oscillatory behavior in certain physical quantities have been reported \cite{maksym1990, wagner1992}. These  oscillatory behaviors are clearly a manifestation of a  dHvA effect type. Since then several authors have been reporting such phenomenon in different systems \cite{miller1995, schwarz2002, terashima2013, romanov2019}. In the following we intend to discuss this subject further.

Therefore, in this article we study the system consisted of two electrons in a single QD with a 3D anisotropic parabolic confining potential under the effect of a magnetic field $B$. We analyze the behavior of some physical quantities as function of B: the singlet and triplet energies, $E_S$ and $E_T$, respectively; the corresponding exchange coupling $J=E_T - E_S$; the two-electron density function; and the spatial spreading of the two electrons in terms of the variance along the $x$-direction.

The present paper is organized as follows. In 
section~ \ref{sec:II}  our theoretical approach is presented. In section~\ref{sec:III}, 
we apply our model for two regimes of  confinement and, in particular, for a $quasi$-2D regime where the effects of the magnetic field are shown in different properties of the system. Finally, in section \ref{sec:IV}, we summarize our conclusions. Throughout
the paper we use atomic units.

\section{\label{sec:II} THEORETICAL APPROACH}

 We want to solve the time independent Schr\"odinger equation for a system of $N$ electrons  submitted to an arbitrary potential $\hat{V}(x,y,z)$ whose Hamiltonian is written as:
\begin{eqnarray}
	\hat H =\sum_j^N\hat{O}_1(\vec r_j)+\sum_j^N\sum_{n<j}^N\hat{O}_2(\vec r_j,\vec r_n),
\label{HO1O2}
\end{eqnarray}
where 
\begin{subequations}
\begin{align}
\hspace{-.5cm} & \hat{O}_1(\vec r_j) = \frac{1}{2m_c}\big[\vec p_j + \vec A(\vec r_j) \big]^2 + g\mu_B\vec S_j \cdot \vec B_j  + \hat{V}(\vec r_j), \\
& \hat{O}_2(\vec r_j,\vec r_n) = \frac{1}{\kappa \vert\vec r_j-\vec r_n\vert},
\end{align}
\label{O1O2}
\end{subequations}
$m_c$ is the effective electronic mass, $e$ is the absolute value of the electron's charge, $g$ is the effective gyromagnetic factor, whose value is $-$0.44 for GaAs, $\mu_B=1/2$ is the Bohr's magneton and $\kappa$ is the relative permittivity or dielectric constant of the QD. It is convenient to choose the gauge 
\begin{eqnarray}
	\vec\nabla \cdot \vec A = 0
\label{divA=0}
\end{eqnarray}
which, in the case of a constant magnetic field 
\begin{eqnarray}
\vec B = B\hat{z}, \nonumber
\end{eqnarray}
yields
\begin{eqnarray}
\vec A(\vec r)= -\frac{1}{2}\vec r \times \vec B = \frac{1}{2}\big( - y\hat{x} + x\hat{y} \big)B.
\label{vecA}
\end{eqnarray}
Detailing the first term of $\hat{O}_1(\vec r_j)$, one gets
\begin{eqnarray}
\hspace{-0.3cm} \frac{1}{2m_c}\big(\vec p_j + e\vec A(\vec r_j) \big)^2 = -\frac{\nabla_j^2}{2m_c} - \frac{i  \vec A \cdot \nabla_j}{m_c} + \frac{\vec A^2}{2m_c},
\label{(p2+eA)^2/2m}
\end{eqnarray}
where we have used Eq.~(\ref{divA=0}).

According to Eq.~(\ref{vecA}) one has
\begin{eqnarray}
- \frac{i}{m_c} \vec A \cdot \nabla = \frac{1}{2m_c}\vec B \cdot \vec L
\label{A.grad}
\end{eqnarray}
and
\begin{eqnarray}
\frac{\vec A^2}{2m_c}= \frac{ B^2}{8m_c}(x^2 + y^2).
\label{e^2A^2/2mc}
\end{eqnarray}

The potential represents the resulting interaction felt by each of the two electrons with all the other particles - electrons and atomic nuclei - that constitute the QD. 
 In the present work we are interested in studying the electronic structure of a system composed of two electrons confined in a anisotropic 3D harmonic quantum dot, whose potential is expressed as
\begin{eqnarray}
\hat{V}(x,y,z) = \frac{m_c}{2}\left(\omega^2_x x^2+\omega^2_y y^2+\omega^2_z z^2\right).
\label{Vq}
\end{eqnarray} 

The electronic properties of confined systems,  such as electrons in a QD, can be obtained due to the flexibility  of our program which can take into account the anisotropy of the potential on the basis employed. In addition, one can  modify the mass of the electron by means of effective electronic mass $m_c$ and/or change the environment in which they evolve via the  dielectric constant $\kappa$ (see Ref.~\cite{carvalho2003}).

 By making use of Eqs.~(\ref{O1O2}), (\ref{vecA}), (\ref{(p2+eA)^2/2m}), (\ref{A.grad}) and (\ref{e^2A^2/2mc}) one can write the operator $\hat O_1(\vec r_j)$ as
\begin{eqnarray}
\hspace{-.5cm}  \hat{O}_1(\vec r_j)& =& -\frac{\nabla_j^2}{2m_c} + \frac{m_c}{2}\left(\omega^2_{x}x_j^2 + \omega^2_{y}y_j^2+\omega^2_z z_j^2\right) \nonumber \\
  && + \frac{B^2}{8m_c}(x_j^2 + y_j^2) + 
    \mu_B \left(\frac{1}{m_c} L_{z j}+ g S_{z j}\right) B \nonumber \\
 & = & -\frac{\nabla_j^2}{2m_c} + \frac{m_c}{2}\left(\Omega^2_{xL}x_j^2 + \Omega^2_{yL}y_j^2+\omega^2_z z_j^2\right) 
  + \nonumber \\
&&   + \mu_B \left(\frac{1}{m_c} L_{zj}+ g S_{zj}\right) B,
\end{eqnarray}
where $ \Omega^2_{xL} = (\omega^2_x +\omega_L^2)$ and  $\omega_L=B/2 m_c$ is  the Larmor frequency. With all these considerations Eq.(\ref{HO1O2}) yields 

\begin{eqnarray}
\hspace{0.8cm} \hat H &=&\sum_j^N\Big[-\frac{\nabla_j^2}{2  m_c}  +\frac{m_c}{ 2 } \Big(\Omega_{xL}^{2} x_j^2  +\Omega_{yL}^{2} y_j^2
\nonumber \\
&& +\omega_{z}^{2} z_j^2 \Big) + \frac{\mu_BB}{m_c} L_{jz}\Big] + \mu_BB  g S_{z}
\nonumber \\
&& + \sum_j^N\sum_{n<j}^N \frac{1}{\kappa \vert\vec r_j-\vec r_n\vert}
\label{hamiltonian}
\end{eqnarray}

\begin{figure}[b]
\centering
\includegraphics[scale=0.45]{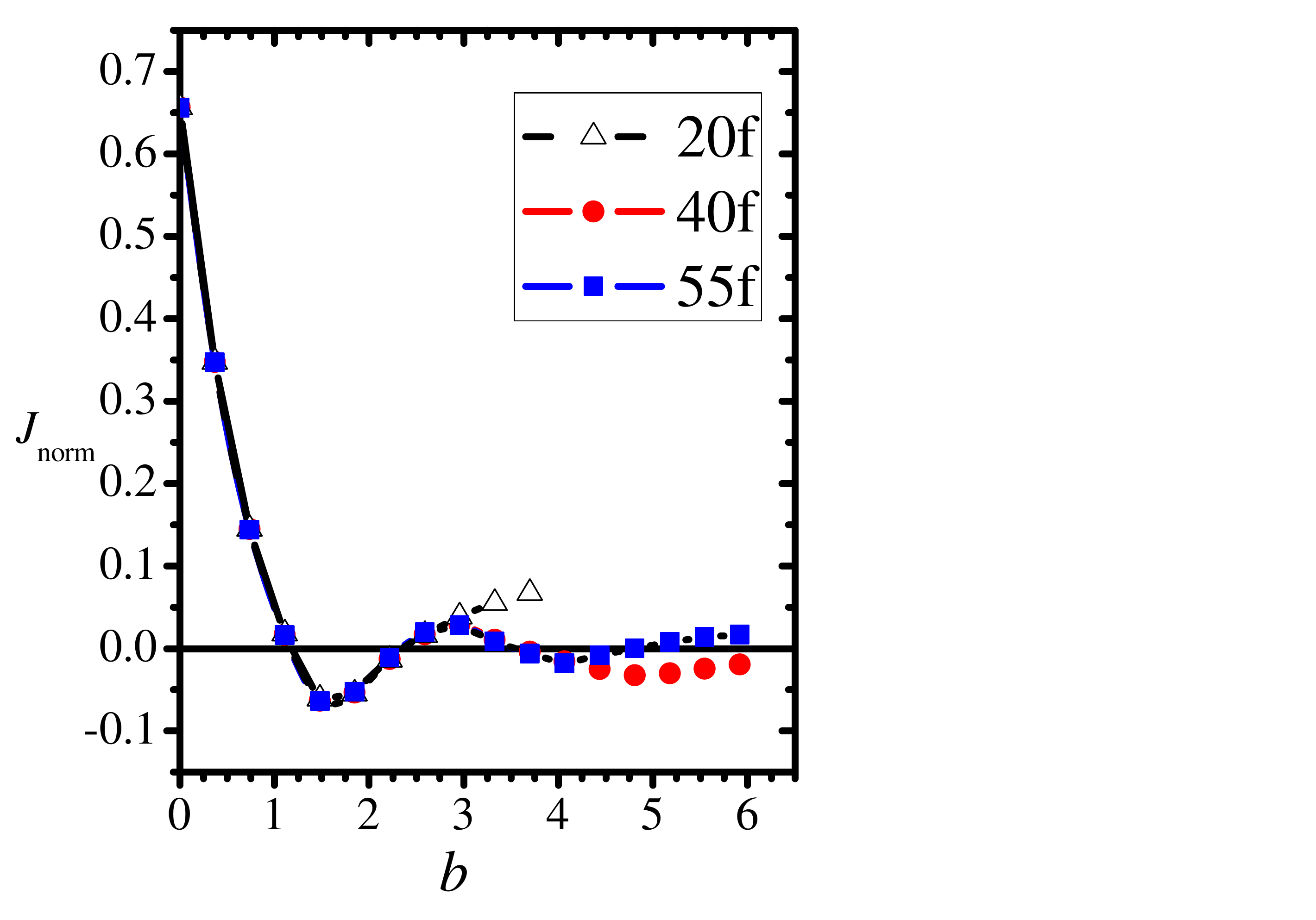}
\caption {$J_{\rm norm} \times b$ for confinement conditions given by $\omega_z=\omega_x=\omega_y=0.000111$ (see the text for details). As highlighted in the legend the open triangles correspond to the results obtained with the basis 20f, whereas the solid circles and squares to the basis 40f and 55f, respectively. }
\label{JxB_wz0,000111}
\end{figure}

The basis employed within the framework of the full configuration interaction (Full-CI) method is similar to the one used in our previous works \cite{olavo2016,maniero2019}, i.e, the functions of the basis are {\it Cartesian anisotropic Gaussian-type orbitals} (c-aniGTO):
\begin{equation}
g^{cart}(\vec{r})=x^{n_x}y^{n_y}z^{n_z}\exp(-\zeta_x x^2 - \zeta_y y^2 - \zeta_z z^2),
\end{equation}
whose exponent $\zeta_i$ ($i=x,~y,~z$) are given by
\begin{equation}
\zeta_i^{(k)} = \frac{m_c\omega_i}{2}\Bigl(\frac{3}{2}\Bigr)^{k -1} ~~(k=1,2),
\end{equation}
where $\zeta_i^{(1)}=m_c\omega_i/2$ when $k=1$ and $\zeta_i^{(2)}=3m_c\omega_i/4$ when $k=2$. In the present case $\omega_i$ is  replaced by $\Omega_{iL}$ for $i=x,y$.  We have employed three different bases: $2s2p2d$, consisting of 20 functions, with 210 (190)  CSFs and 400 (190) determinants for the singlet (triplet) states;  $2s2p2d2f$, consisting of 40 functions, with 820 (780)  CSFs and 1600 (780) determinants for the singlet (triplet) states; and $2s2p2d2f1g$, consisting of 55 functions, with 1485 (1485)  CSFs and 2916 (1485) determinants for the singlet (triplet) states. For the sake of simplicity, from now on we shall refer them as 20f, 40f, and 55f, respectively.

Therefore, the solution of the Sch\"odinger equation $\hat H\Phi=E\Phi$ is written as a linear combination of configuration state functions (CSFs),
\begin{equation}
\Phi=\sum_{n=1}^{N_\textrm{CSF}}C_n^{\textrm{CSF}}\Psi_n^{\textrm{CSF}},
\label{Phi}
\end{equation}
where $N_\textrm{CSF}$ is the number of CSFs, $\Psi_n^{\textrm{CSF}}$ is the $n$th CSF, and $C_n^\textrm{CSF}$ is its corresponding coefficient. On the other hand, a CSF is constituted of Slater determinants, {\em i.e.},
\begin{eqnarray}
\Psi_n^\textrm{CSF}=\sum_{m=1}^{\textrm{Ndet}_n}C_{m}^\textrm{det}\textrm{det}_{n,m},
\end{eqnarray}
where $\textrm{det}_{n,m}$ is the \emph{mth} determinant of the \emph{nth} CSF.

In Appendix A we show the expressions of the angular momentum integrals that were implemented in our numerical code. They depend only of the overlap integrals which have already been used in our previous works.

\begin{figure}[b]
\centering
\includegraphics[scale=0.45]{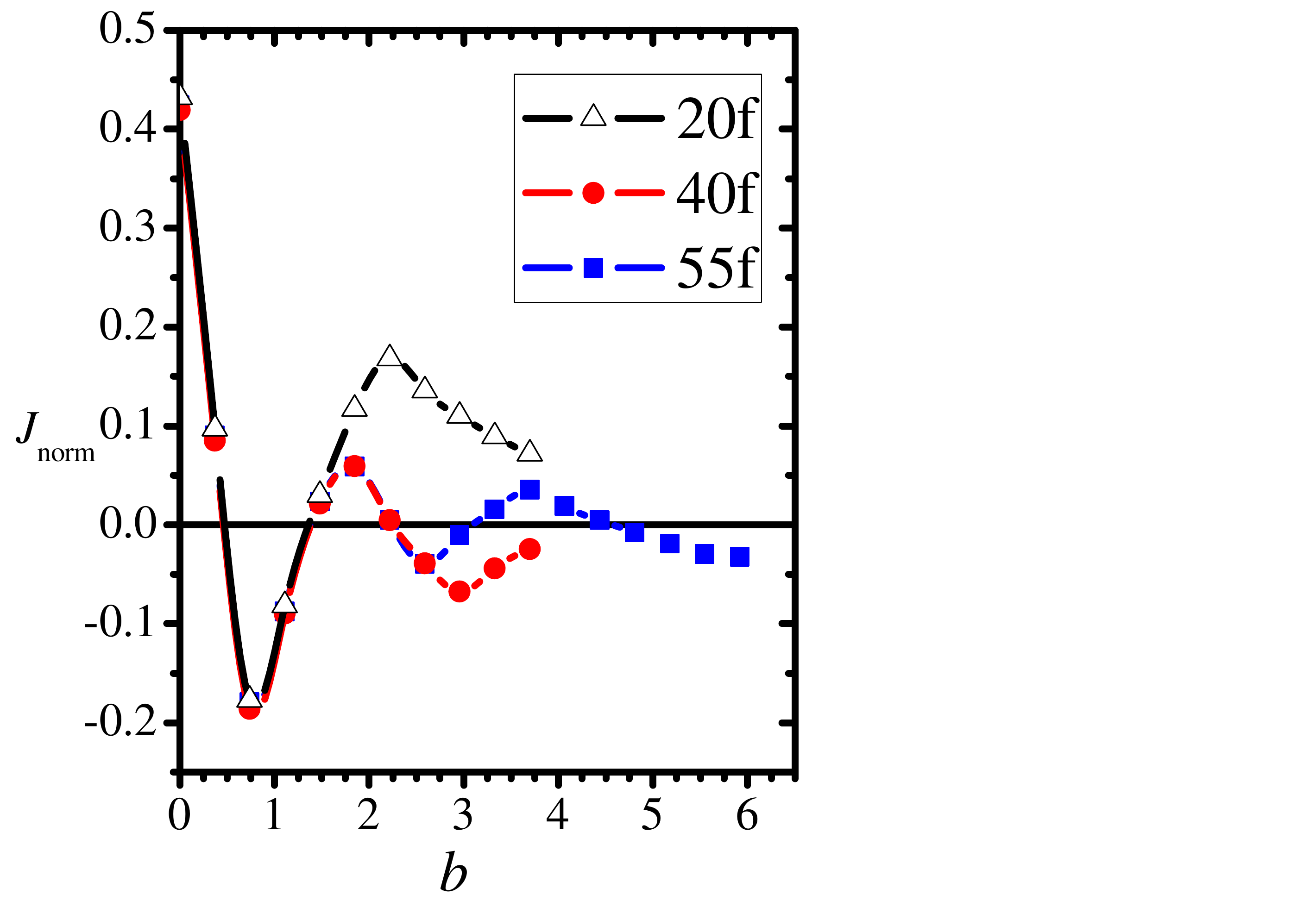}
\caption {$J_{\rm norm} \times b$ for confinement conditions where $\omega_z \gg \omega_x, \omega_y$: $\omega_z=1.11$ and $\omega_x=\omega_y=0.000111$. As highlighted in the legend the open triangles correspond to the results obtained with the basis 20f, whereas the solid circles and squares to the basis 40f and 55f, respectively.}
\label{JxB_wz1.11}
\end{figure}

\section{\label{sec:III} RESULTS AND DISCUSSION}

For a proper  parameterization of our problem, we shall express the magnetic field in terms of a characteristic magnetic field $B_c=2m_c\omega_x$, as well as the energy in terms of $\omega_x$. In all calculations, we consider $\omega_x=\omega_y=0.000111$, while we assume the values 0.000111 and 1.11 for $\omega_z$. For $\omega_x=0.000111$ one has an energy of 3 meV and $B_c \approx 1.49\times 10^{-5}$ ( $\approx 3.5$ T) .  In what follows, we shall consider $S_z=0$.

In  Figs.~\ref{JxB_wz0,000111} and~\ref{JxB_wz1.11} we display the behavior of $J_{\rm norm}$ ($J_{norm}=J/ \omega_x$) as function of the normalized field $b$ ($b=B/B_c$) for $\omega_z=0.000111$ and $\omega_z=1.11$, respectively. The first thing we notice in the behavior of $J_{\rm norm}$ is its oscillatory comportment around zero, meaning a successive change of the ground state between the lowest singlet ($J_{\rm norm} > 0$) and the lowest triplet states ( $J_{\rm norm} < 0$). As it is well known, the accuracy of the results depends largely on the atomic basis used.

For the isotropic (spherically symmetric) case, one observes from Fig.~\ref{JxB_wz0,000111} that, until $b\approx 3$, the result obtained with the basis 20f is in good accordance with the ones obtained with the bases 40f and 55f. Beyond this value, it diverges from the two others. Therefore, one can see that the basis 20f describes well the first oscillation of $J_{\rm norm}$ which occurs in the range of $b$ from 0 to approximately 3. In addition, the results for $J_{\rm norm}$ predicted by the computation with the bases 40f and 55f remain in good accordance until $b\approx 4$, where they split, but yet remaining close till larger values of $b$ ($\approx$ 6). Hence, the introduction of functions with $l=3$ in the basis set, what happens when one passes from basis 20f to 40f, allows one to describe well the first part of the second oscillation which occurs approximately in the interval of $b$ from 3 to 4. But the second part of it, from $b=4$ to 6, the basis 40f is not sufficiently accurate to describe it. 

It is worth saying that we can be confident of a result when it is obtained with different basis sets. This is what happens in the case of $J_{\rm norm}$ with $b$ from 0 to $\approx 3$ or till $\approx 4$. However, when different predictions arise from the use of different bases, we know that the better outcome is the one obtained with the largest basis, but we cannot say how good it is.

In Fig.~\ref{JxB_wz1.11} we observe a similar behavior to that seen in the previous figure. In this case, however, the confining potential is no longer spherically symmetric, as happens in the former. Now, the confinement strength in the $z$-axis is much stronger than in the other two directions ($\omega_z \gg \omega_x = \omega_y$) and, thus, one can consider it as a $quasi$-2D case. Therefore, although $J_{\rm norm}$ presents a similar behavior, one can notice that it oscillates faster as a function of $b$. Besides, the concordance between the results obtained with different basis sets become worse;  the results of the three sets of bases agree only up to $b \approx 1.5$ and the concordance between bases 40f and 55f remains just up to $b \approx 2.5$. 

 The results from Figs.~\ref{JxB_wz0,000111} and~\ref{JxB_wz1.11} show the need of having atomic basis functions with high $l$ in the CI calculation to properly describe the oscillations between singlet and triplet states. Additionally, they indicate that the fundamental state of a two-electron QD  alternates between singlet and triplet states, more precisely between states with  $l=0$ (singlet), $l=1$ (triplet), $l=2$ (singlet), $l=3$ (triplet), and so on, in accordance  with previous results in the literature~\cite{wagner1992, dineykhan97, ban2rjee08}. Other interesting observation is that the $J_{\rm norm}$ oscillations occur for larger $b$ values in the 3D case ($w_z=0.000111$) than in the $quasi$-2D one ($w_z=1.11$). This behavior is also observed in the study of the chemical potential and the addition energy at Ref.~\cite{poszwa16}.

From now on we shall focus our attention in the quasi-2D  case ($w_z=1.11$). All physical quantities, that can be experimentally observed, are connected with the electrons' spatial distribution. Thus, let us turn our attention to the behavior of the localization of an electron along an axis, say the $x$-axis, expressed in terms of its root-mean-square position ($\sigma_x = \sqrt{\langle x^2 \rangle}$). Hence, one has
\begin{eqnarray}
\langle x^2 \rangle=\int dx_1dx_2 ~\rho(x_1,x_2)~ x_1^2, 
\label{<x^2>}
\end{eqnarray}
where  
\begin{eqnarray}
\rho(x_1,x_2)=\int d\beta_1d\beta_2 dy_1dy_2 dz_1dz_2  \vert\Phi\vert^2, 
\label{rhox1x2}
\end{eqnarray}
is the two-electron localization distribution along the $x$-axis, $\Phi=\Phi(\vec r_1, \vec r_2,\beta_1,\beta_2)$ is the solution of Eq.~(\ref{hamiltonian}), $\beta_1$ and $\beta_2$ stand for the spin coordinates of the two electrons, and $\vec r_1=(x_1,y_1,z_1)$ and $\vec r_2=(x_2,y_2,z_2)$ their spatial coordinates. In order to compute Eq.~(\ref{rhox1x2}) we have adopted the following approximation: in the expansion of $\Phi$, Eq.~(\ref{Phi}), we have only taken into account the CSFs, whose coefficient has modulus greater than 0.09.

In fact, both quantities, $\sigma_x$ and $\rho(x_1,x_2)$, give interesting information on the behavior of the electrons inside the QD in the presence of the magnetic field and we shall discuss them in next.

\begin{figure}[h]
\centering
\includegraphics[scale=0.45]{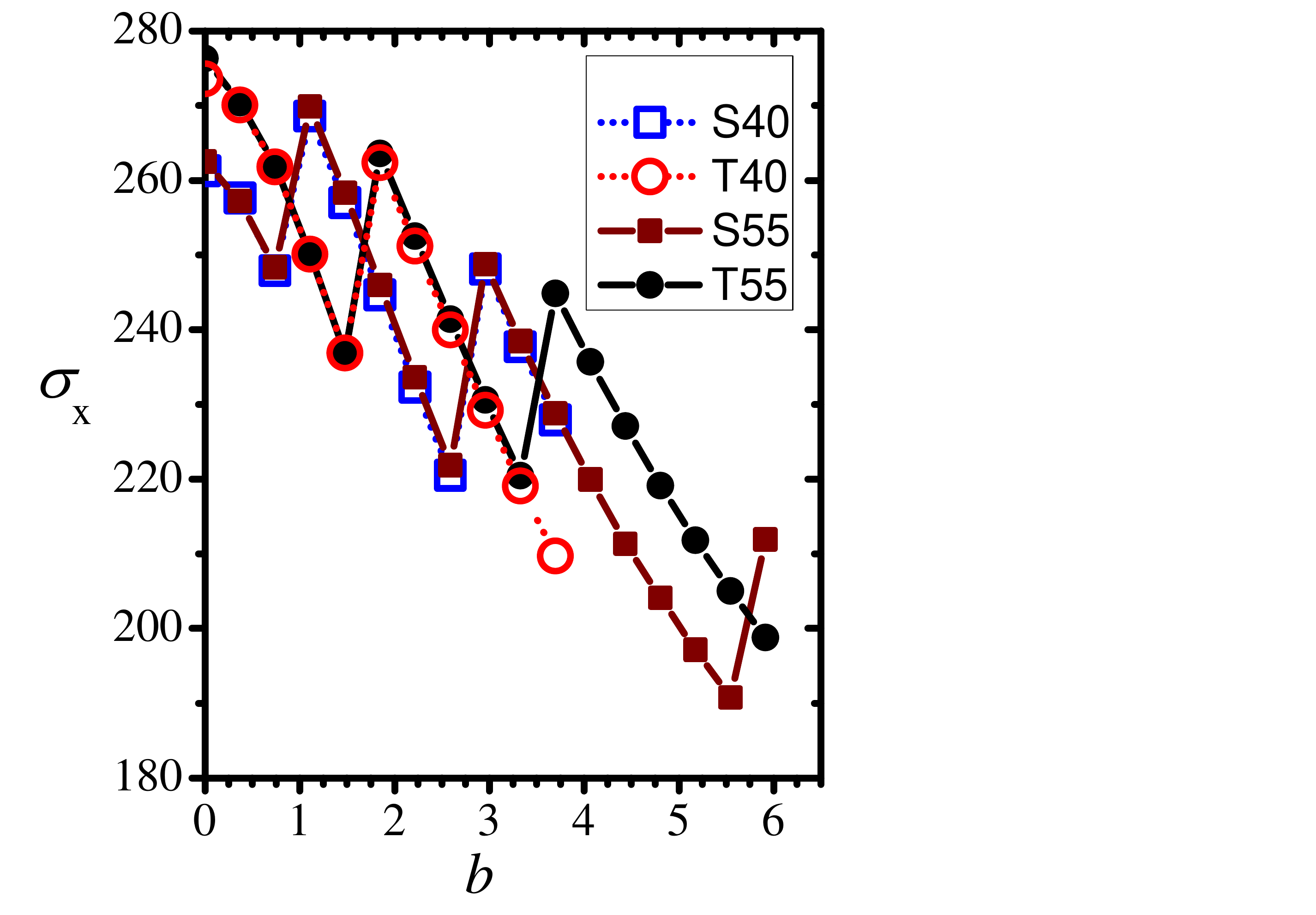}
\caption {$\sigma_x$ as a function of $b$ for the singlet and triplet states. The dotted lines and open symbols correspond to the results obtained with the basis 40f, and the solid lines and solid symbols correspond to the basis 55f. The legend details the singlet and triplet curves; for instance, S40 means singlet and basis 40f, and so on.}
\label{sqrtx2}
\end{figure}

Thus, in Fig.~\ref{sqrtx2} we display the behavior of $\sigma_x$, for the singlet and triplet states, as a function of the magnetic field, computed by using the bases 40f and 55f. The first thing to notice is that the two bases are in good agreement up to $b \approx 3.3$; up to this point $\sigma_x$ develops more than two oscillations, for the singlet state, and, for the triplet, the second oscillation is almost complete. At $b \approx 3.7$ the results for the triplet state obtained by the two bases no longer agree, whereas for the singlet they still remain in accordance. Besides the oscillatory comportment, present in  $J_{\rm norm}$ and also here, we observe another interesting aspect as well. In Fig.~\ref{JxB_wz1.11} one has the extremes of $J_{\rm norm}$, its minimums and maximums, occurring approximately at $b = 0.75, 1.8, 2.6, 3.7$ if one follows the curve of the 55f result. By observing Fig.~\ref{sqrtx2} one can see that, at the vicinity of these values of $b$, the curves of $\sigma_x$ for the singlet and triplet states cross. At first we expected to see a correlation between the points where $J_{\rm norm}=0$ and the ones of the crossing of $\sigma_x$ for the singlet and triplet. However, if there is any type of correlation, it seems to be with $dJ_{\rm norm}/db=0$. This is an issue to be addressed in a future work.

\begin{figure*}[pt]
\vspace{-.85 cm}
\centering
\includegraphics[width=\textwidth]{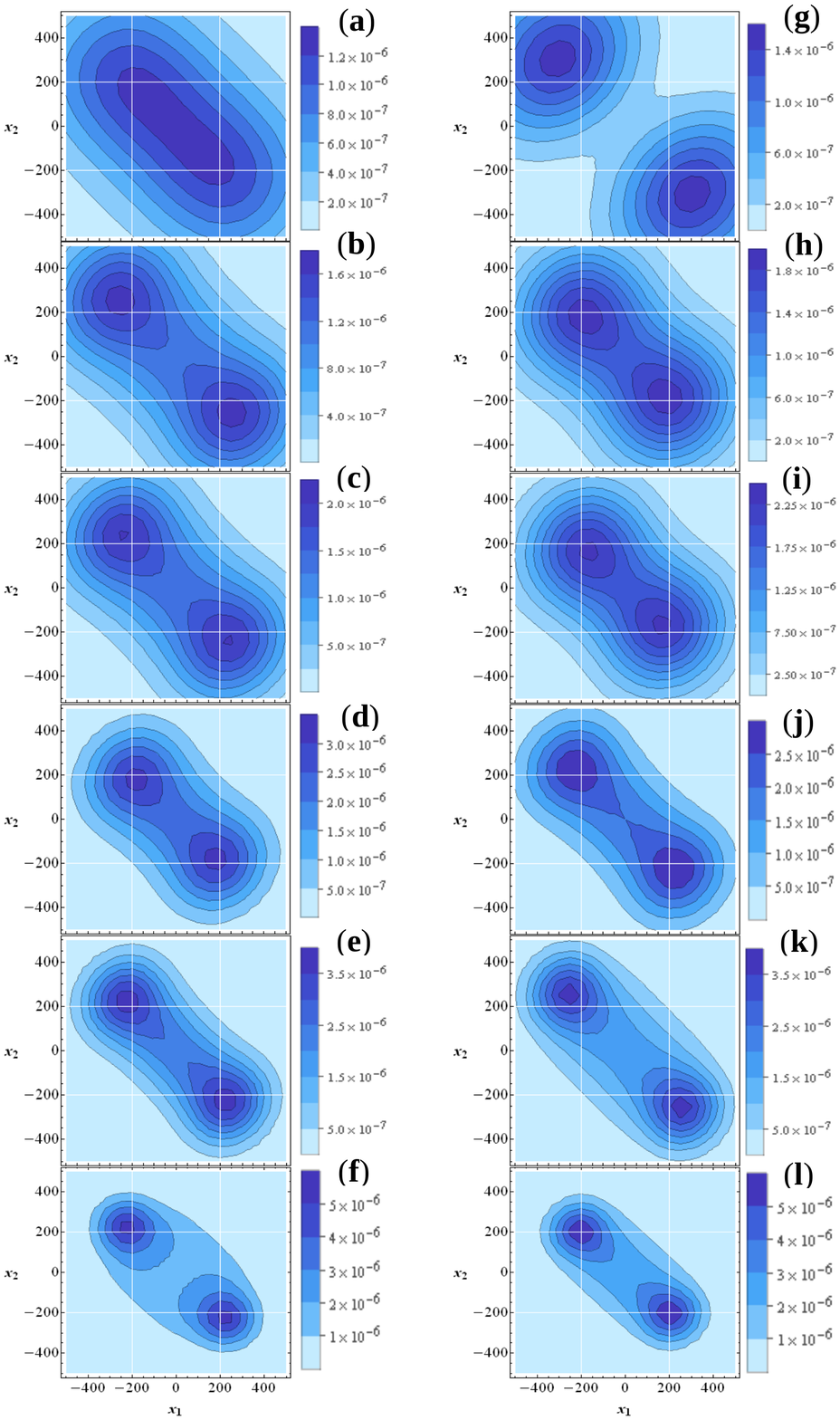}
\vspace{-1.5 cm}
\caption {Contour map of $\rho(x_1,x_2)$ -- the spacial distribution of the two electrons along the $x$-axis -- for the singlet and triplet states. The left panels (a -- f) corresponds to the singlet state and the right ones (g -- l) to triplet. The confinement conditions are the same of Fig.~\ref{JxB_wz1.11} and corresponds to a $quasi$-2D QD. From top (a, g) to botton (f, l) $b=0, 1.11, 1.48, 2.59, 3.70, 5.92$, respectively.}
\label{contour_plot}
\end{figure*}

Now, let us analyze the behavior of $\rho(x_1,x_2)$.  In Fig.~\ref{contour_plot} are displayed the contour maps of $\rho(x_1,x_2)$ for the singlet and triplet states under different values of the magnetic field; the confinement strength conditions are the same of Fig.~\ref{JxB_wz1.11},{\it i.e.}, $\omega_z=1.11$ and $\omega_x=\omega_y=0.000111$. The contours are plotted in the phase-space $x_1 \times x_2$ that corresponds to the positions of electron 1 and 2, respectively, along the $x-$axis; since the electrons are indistinguishable, one finds a reflection symmetry with respect to the diagonal line $x_2=x_1$ corresponding to the exchange of the electrons. 

The situation depicted by panel (a) shows that the electrons, in the singlet state for $b=0$, are spread throughout the phase-space, being more concentrated along the line $x_2=-x_1$ and its surroundings, mainly in the interval $x \in (-200, 200)$. On the other hand, the panel (g) shows the electrons in the triplet state for $b=0$  concentrated at $x = \pm 300$. From these two figures, one would expect that the interaction energy $1/(\kappa \vert\vec r^{\,\prime}_i-\vec r^{\,\prime}_j\vert)$ between the two electrons would make the singlet energy higher than that of the triplet, leading to $J < 0$. However, Fig.~\ref{JxB_wz1.11} shows us the opposite; we must remember that the distributions along the two others directions, in particular the $y$-direction  -- since it is a $quasi$-2D system --, have also to be considered and, in the present case, they are responsible to make $J > 0$, as shown in Fig.~\ref{JxB_wz1.11}. The objective of Fig.~\ref{contour_plot} is indeed to give an idea of how the magnetic field affects the spatial distribution of the electrons. 

Now, going ahead on the others panels of Fig.~\ref{contour_plot}, we observe that for $b=1.11$ the singlet (panel (b)) has concentrated at $x \approx \pm 250$, whereas the triplet (panel (h)) has their peaks moved from $x = \pm 300$ to $x \approx \pm 175$. For $b=1.48$ one finds the singlet (panel (c)) and the triplet (panel (i)) with their peaks approximately at the same position, $x \approx \pm 250$ and $x \approx \pm 175$, respectively. However, their electron distributions become narrower; one can see this from the scale displayed on the right of the panels, where the top values in both cases are larger. For $b=2.59$ the singlet distribution has its peak moved backward from $x \approx \pm 250$ to $x \approx \pm 175$ (panel d) and the triplet has its peak distribution moved forward from $x \approx \pm 175$ to $x \approx \pm 225$ (panel j).  For $b=3.70$ the singlet peak distribution has gone back from  $x \approx \pm 175$ to $x \approx \pm 225$ (panel e) and the triplet one has gone further from $x \approx \pm 225$ into $x \approx \pm 250$ (panel k). Finally in the last two panels, (f) and (l), corresponding to $b=5.92$, one can see the singlet peak approximately at the same position, $x \approx \pm 225$ and the triplet peak backward at $x \approx \pm 200$. Therefore, there is a back and forth movement around $x \approx \pm 200$ of the electron distributions in the singlet and triplet states as the magnetic field increases, and this movement is accompanied by a narrowing of the distribution. Observe that these movements of both distributions are not synchronized, leading to a complicated oscillatory behavior of physical quantities, such as $J_{\rm norm}$, as a function of the magnetic field.

 \section{\label{sec:IV} CONCLUSIONS} 

In this paper we have studied a two-electron QD in the presence of a magnetic field. Our focus was the properties of the singlet and triplet lowest states, such as their energy, the exchange coupling and other ones. We have analyzed their behavior as a function of the magnetic field intensity under two different confinement conditions. In one case we have considered a 3D isotropic confinement potential and the other, a $quasi$-2D one. In both cases the potential was parabolic along the three axes, but in the latter the strength along the $z$-direction was taken much greater than in the other two directions. Our analyze relied on numerical results obtained by using the Full-CI method with Cartesian anisotropic Gaussian-type orbitals as basis set.  The confidence of the results have been discussed in terms of the convergence of the results as a function of the size of the basis set. In addition, we have reported an oscillatory motion of the physical quantities as a function of the magnetic field. This motion obtained from our computation resembles a phenomenon found in the literature known as the dHvA  effect. What is quite amazing in this fact is that a semiconductor QD, with more than one electron, seems to satisfy the condition of having their inner electrons partially free, or partially confined, as predicted by de Haas and van Alphen 90 years ago. The understanding of the physics behind the interplay of a confinement potential together with the Coulomb interaction and the quantum requirements fulfilled by the electrons in the presence of a magnetic field is a challenging task; we believe that QD with more than one electron is a promising system to help to achieve it.

\begin{acknowledgments}

This work was partially supported by the Brazilian agencies CNPq,  CAPES, FAPESB and FAPERJ.

\end{acknowledgments}
\appendix
\section {Angular momentum integrals}
\label{ang-mom-int}

The angular momentum  integral is given by
\begin{eqnarray*}
 I^{L_Z}_{\mu \nu}=
	\int_{-\infty}^\infty d\vec r 
g_\mu(\vec r,\vec{\zeta})
\left[-i \left(x\frac{ \partial}{\partial y} -y\frac{ \partial}{\partial x}\right)\right]
g_\nu(\vec r,\vec{\eta}),
\end{eqnarray*}

Defining the overlap integral $S_x(n_x,m_x)$ as

\begin{eqnarray}
S_x(n_x,m_x)= \int_{-\infty}^\infty dx
x^{n_x+m_x}
	e^{-(\zeta_x+\eta_x)x^2}\,
\label{eqSx}
\end{eqnarray}
and similarly for $S_y(n_y,m_y)$ e $S_z(n_z,m_z)$, we obtain 
%
\begin{eqnarray}
	&&I^{L_Z}_{\mu \nu}= -i S_z(n_z,m_z)\Big\{I^P_x(n_x,m_x)\Big[-2\zeta_y S_y(n_y+1,m_y)
\nonumber \\	
&&-2\eta_y S_y(n_y,m_y+1)+ n_y S_y(n_y-1,m_y) + m_y S_y(n_y,m_y-1) \Big] 
\nonumber \\
&& - I^P_y(n_y,m_y)\Big[-2\zeta_x S_x(n_x+1,m_x)-2\eta_x S_x(n_x,m_x+1) 
\nonumber \\ 
&&+ n_x S_x(n_x-1,m_x)  + m_x S_x(n_x,m_x-1) \Big] \Big\}
\end{eqnarray}
\noindent where  $I_x^{Px}$ has the following form
\begin{eqnarray}
	I_x^{P}= \int_{-\infty}^\infty dx x^{n_x+m_x+1} e^{-(\zeta_x+\eta_x)x^2} 
\end{eqnarray}
\begin{eqnarray}
I_x^{P}=  S_x(n_x,m_x+1)=S_x(n_x+1,m_x)
\end{eqnarray}
which shows that our result depends solely upon the overlap integrals whose values are known (see Ref.~\cite{Ho12}) and similarly for $I_y^P$.

\bibliographystyle{unsrt}
\bibliography{RefQDots}

\end{document}